\documentclass[manuscript]{aastex}

\shorttitle{Transverse component of the magnetic field}
\shortauthors{Danilovic et al.}

\begin{document}


\title{Transverse component of the magnetic field in the solar photosphere observed by Sunrise}

\author{\textsc{
S.~Danilovic,$^{1}$ B.~Beeck,$^{1}$ A.~Pietarila,$^{1}$
M.~Sch\"ussler,$^{1}$ S.~K.~Solanki,$^{1,7}$ V.~Mart\'inez
Pillet,$^{2}$ J.~A.~Bonet,$^{2}$ J.~C.~del~Toro~Iniesta,$^{4}$
V.~Domingo,$^{5}$ P.~Barthol,$^{1}$ T.~Berkefeld,$^{3}$
A.~Gandorfer,$^{1}$ M.~Kn\"olker,$^{6}$ W.~Schmidt,$^{3}$ \&
A.~M.~Title$^{8}$ }}

\affil{
$^{1}$Max-Planck-Institut f\"ur Sonnensystemforschung, Max-Planck-Str. 2, 37191 Katlenburg-Lindau, Germany.\\
$^{2}$Instituto de Astrof\'{\i}sica de Canarias, C/Via L\'actea s/n, 38200 La Laguna, Tenerife, Spain.\\
$^{3}$Kiepenheuer-Institut f\"ur Sonnenphysik, Sch\"oneckstr. 6, 79104 Freiburg, Germany.\\
$^{4}$Instituto de Astrof\'{\i}sica de Andaluc\'{\i}a (CSIC), Apartado de Correos 3004, 18080 Granada, Spain.\\
$^{5}$Grupo de Astronom\'{\i}a y Ciencias del Espacio, Universidad de Valencia, 46980 Paterna, Valencia, Spain.\\
$^{6}$High Altitude Observatory, National Center for Atmospheric Research, P.O. Box 3000, Boulder CO 80307-3000, USA.\\
$^{7}$School of Space Research, Kyung Hee University, Yongin, Gyeonggi, 446-701, Korea.\\
$^{8}$Lockheed Martin Solar and Astrophysics Laboratory, Bldg.
252, 3251 Hanover Street, Palo Alto, CA 94304, USA. }

\email{danilovic@mps.mpg.de}

\begin{abstract}
We present the first observations of the transverse component of
photospheric magnetic field acquired by the imaging magnetograph
{\sc Sunrise}/IMaX. Using an automated detection method, we obtain
statistical properties of 4536 features with significant linear
polarization signal. We obtain a rate of occurrence of
$7\cdot10^{-4}$~s$^{-1}$~arcsec$^{-2}$, which is $1-2$ orders of
magnitude larger than values reported by previous studies. We show
that these features have no characteristic size or lifetime. They
appear preferentially at granule boundaries with most of them
being caught in downflow lanes at some point. Only a small
percentage are entirely and constantly embedded in upflows
($16\%$) or downflows ($8\%$).
\end{abstract}


\section{Introduction}

Recent observations with high spatial resolution and polarimetric
sensitivity revealed that quiet photospheric regions contain a
large amount of horizontal magnetic field
\citep{Orozco:etal:2007a,Orozco:etal:2007b,Lites:etal:2008}. The
size of the horizontal field patches varies from less than one to
a few arcsec
\citep{Lites:etal:1996,DePontieu:2002,Marian:etal:2007,Harvey:etal:2007,Ishikawa:etal:2008,Jin:etal:2009}.
Those with sizes comparable to the average size of the granular
pattern are very dynamic
\citep{Ishikawa:etal:2008,Ishikawa:Tsuneta:2009,Jin:etal:2009}.
Such Horizontal Internetwork Fields (HIF) appear in internetwork
as well as in plage regions with no significant difference in the
rate of occurrence. Their lifetimes range from a minute to about
ten minutes, comparable to the lifetime of granules. Some of them
are recognized to be loop-like structures emerging cospatially
with granules. They appear first inside the granule, then move to
the intergranular lanes where they disappear
\citep{Centeno:etal:2007,Gomory:etal:2010}. Around $23\%$ of such
loop-like features rise and thus may contribute to the heating of
higher atmospheric layers \citep{Marian:Luis:2009}. On the other
hand, some HIF are associated with downflows
\citep{Kubo:etal:2010}.

MHD simulations show that horizontal magnetic field may appear
during flux cancellations \citep{Stein:Nordlund:2006} or flux
emergence over single or multiple granules
\citep{Steiner:etal:2008,Cheung:etal:2007}. Additionally, a
significant amount of small-scale horizontal field is possibly
produced through local dynamo action
\citep{Schuessler:Voegler:2008}. To estimate what fraction of HIF
has its origin in each of these processes could, however, be
challenging since their observable signature may be similar.

Previous studies of HIF were based on slit observations of
selected features that appear as single events, mostly associated
with upflows. Here we use the first imaging observations obtained
with the Imaging Magnetograph eXperiment
\citep[IMaX,][]{Valentin:etal:2004,Valentin:etal:2010} onboard
{\sc Sunrise}, a balloon-borne solar observatory
\citep{Barthol:etal:2010,Solanki:etal:2010,Berkefeld:etal:2010,Gandorfer:etal:2010},
to obtain statistical properties of HIF. Mounted on a 1-m aperture
telescope, IMaX provides two-dimensional maps of the vector
magnetic field with exceptional spatial and temporal resolution.
Using these data, we study all structures that show significant
linear polarization signal in the selected quiet Sun time series.
We examine their properties, in particular their connection with
the velocity field.

\begin{figure*}
\includegraphics*[scale=.3]{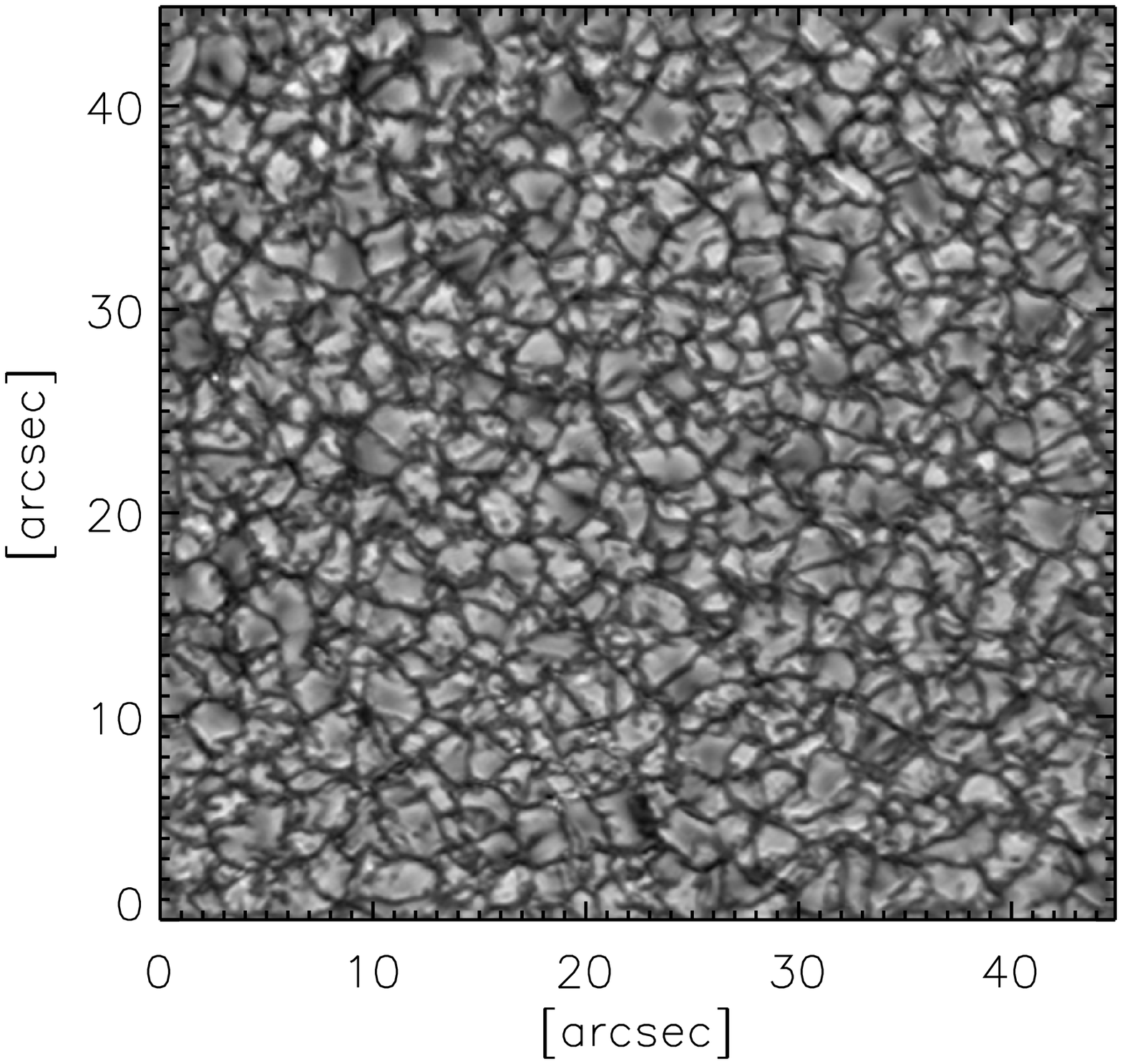}
\includegraphics*[scale=.3]{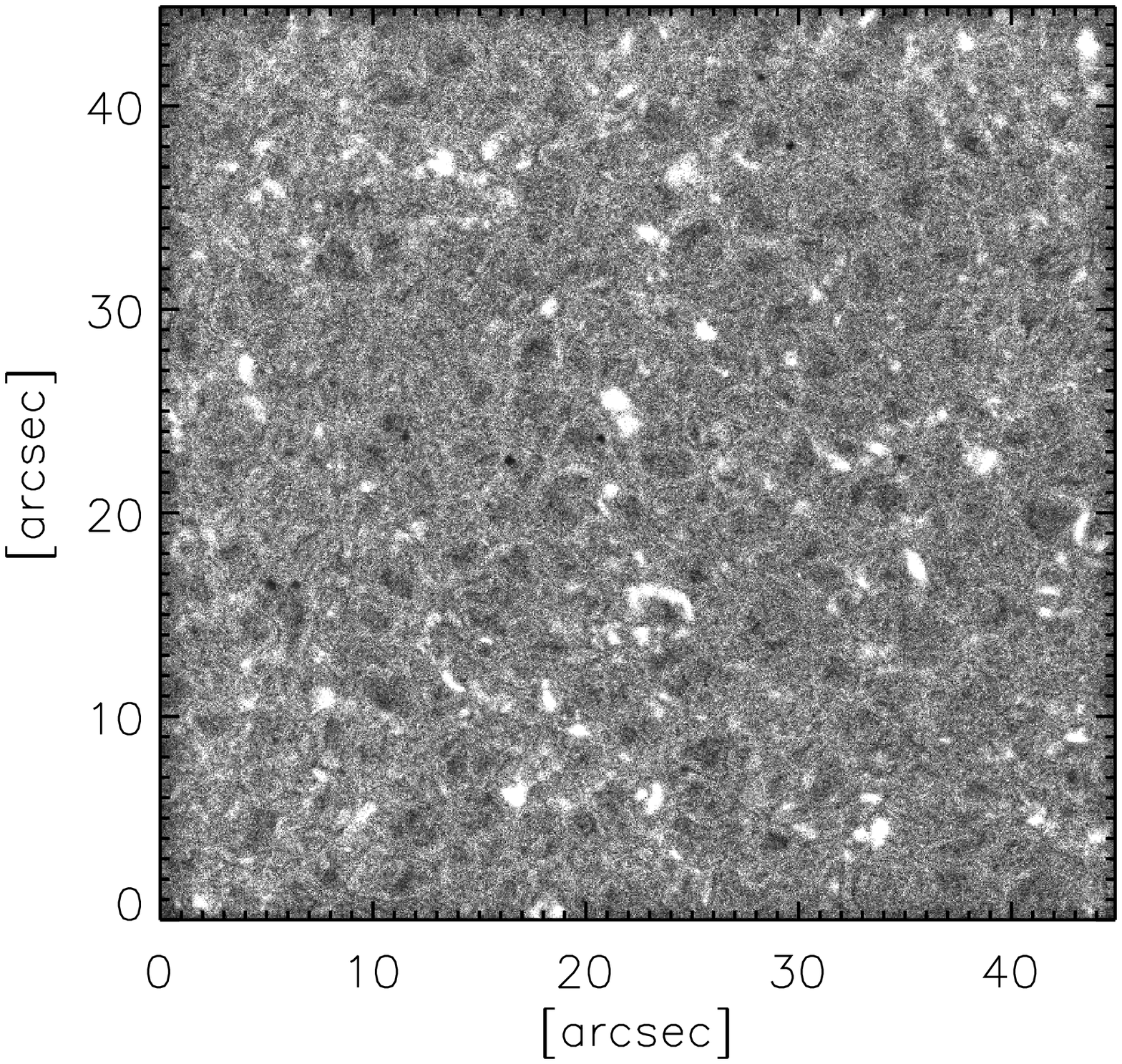}
\includegraphics*[scale=.3]{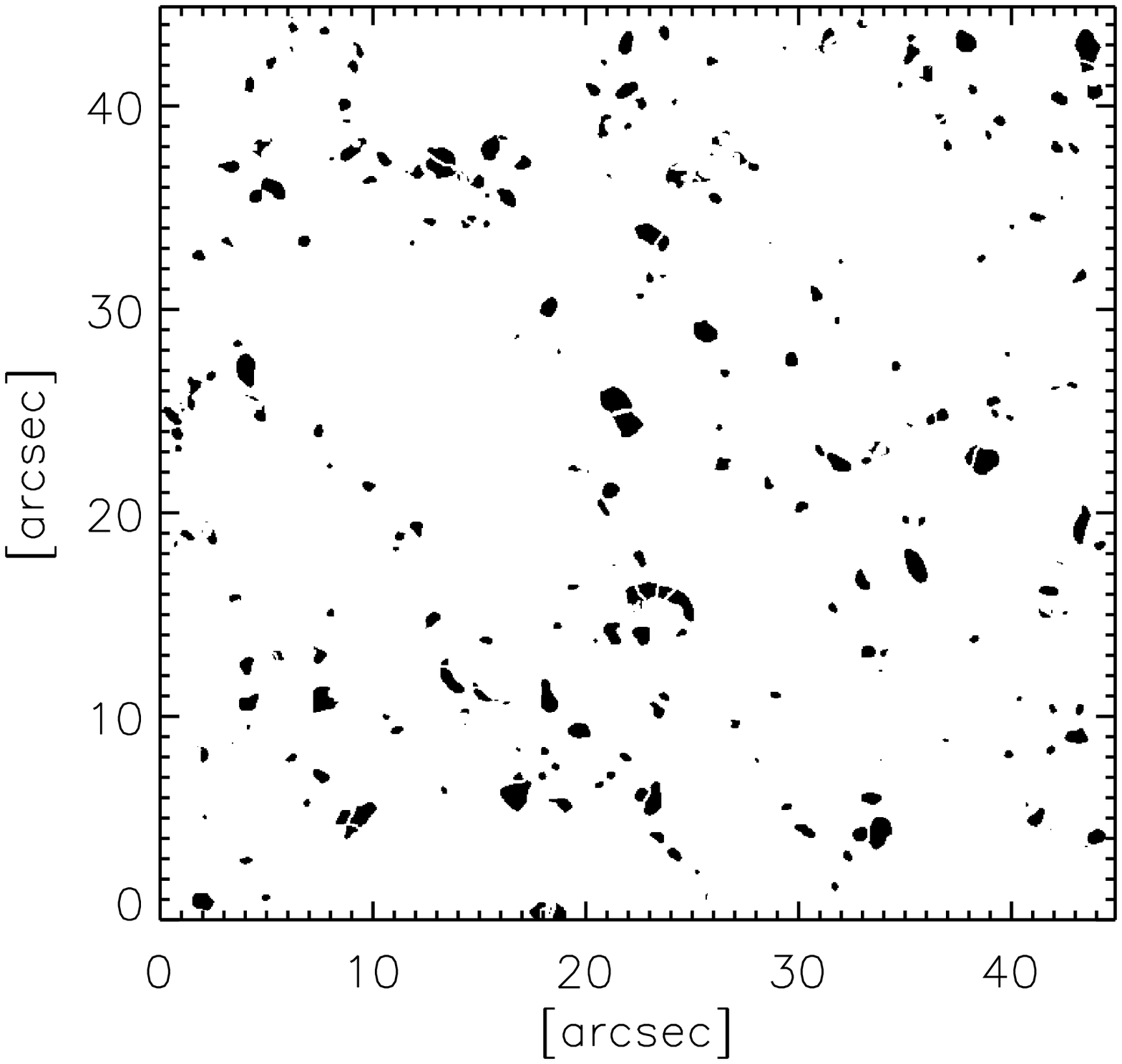}
\caption{An example of a continuum intensity map (left), linear
polarization map (middle) and the corresponding map of the
features detected with the MLT algorithm (right). The temporal
evolution of these maps (01:30:54-02:02:29~UT) is shown in the
movie provided as online material
(http://www.mps.mpg.de/data/outgoing/danilovic/hif/).
\label{fig:masks}}
\end{figure*}

\begin{figure*}
\centering
\includegraphics*[angle=90,scale=.17]{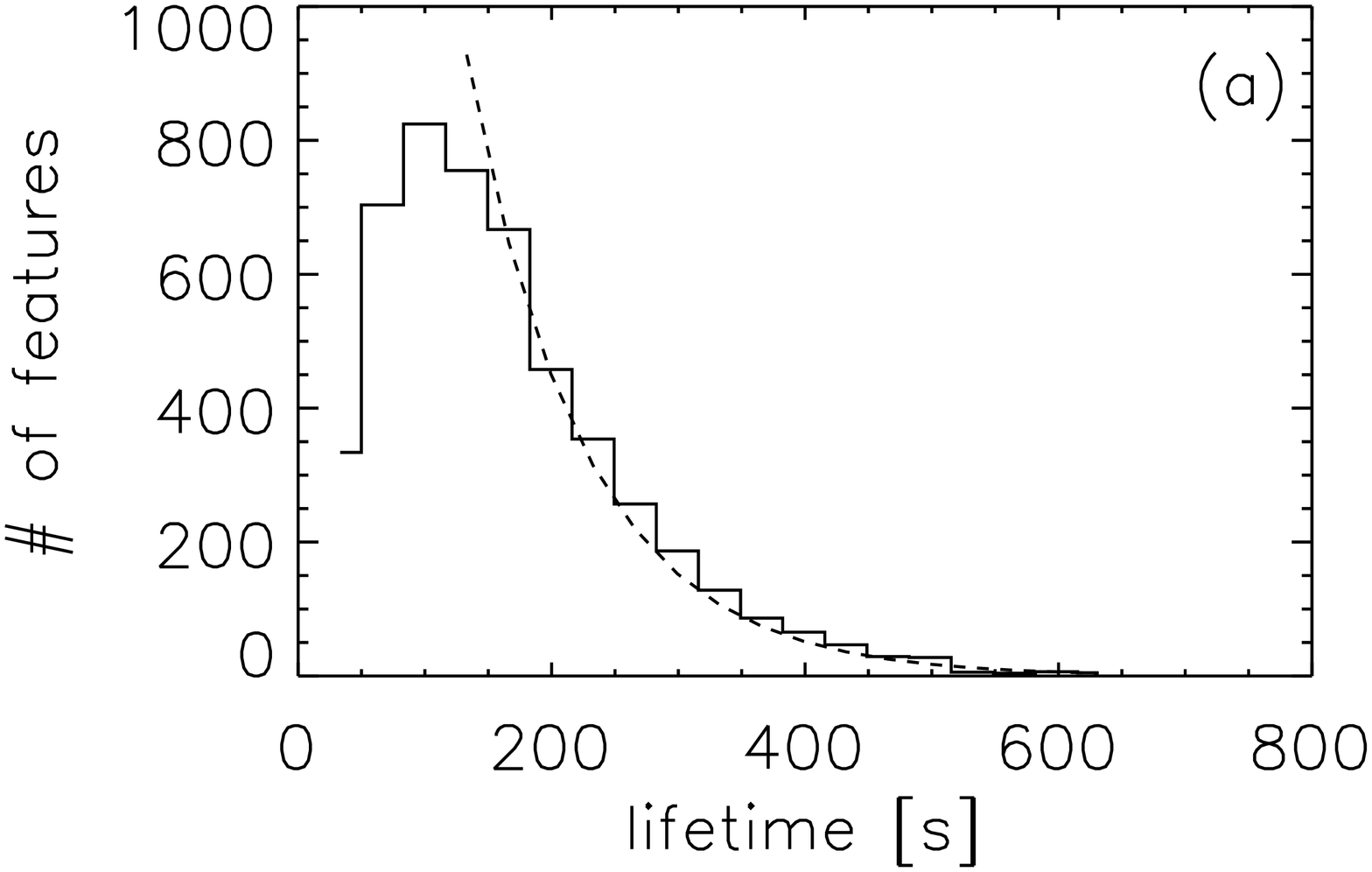}
\includegraphics*[angle=90,scale=.17]{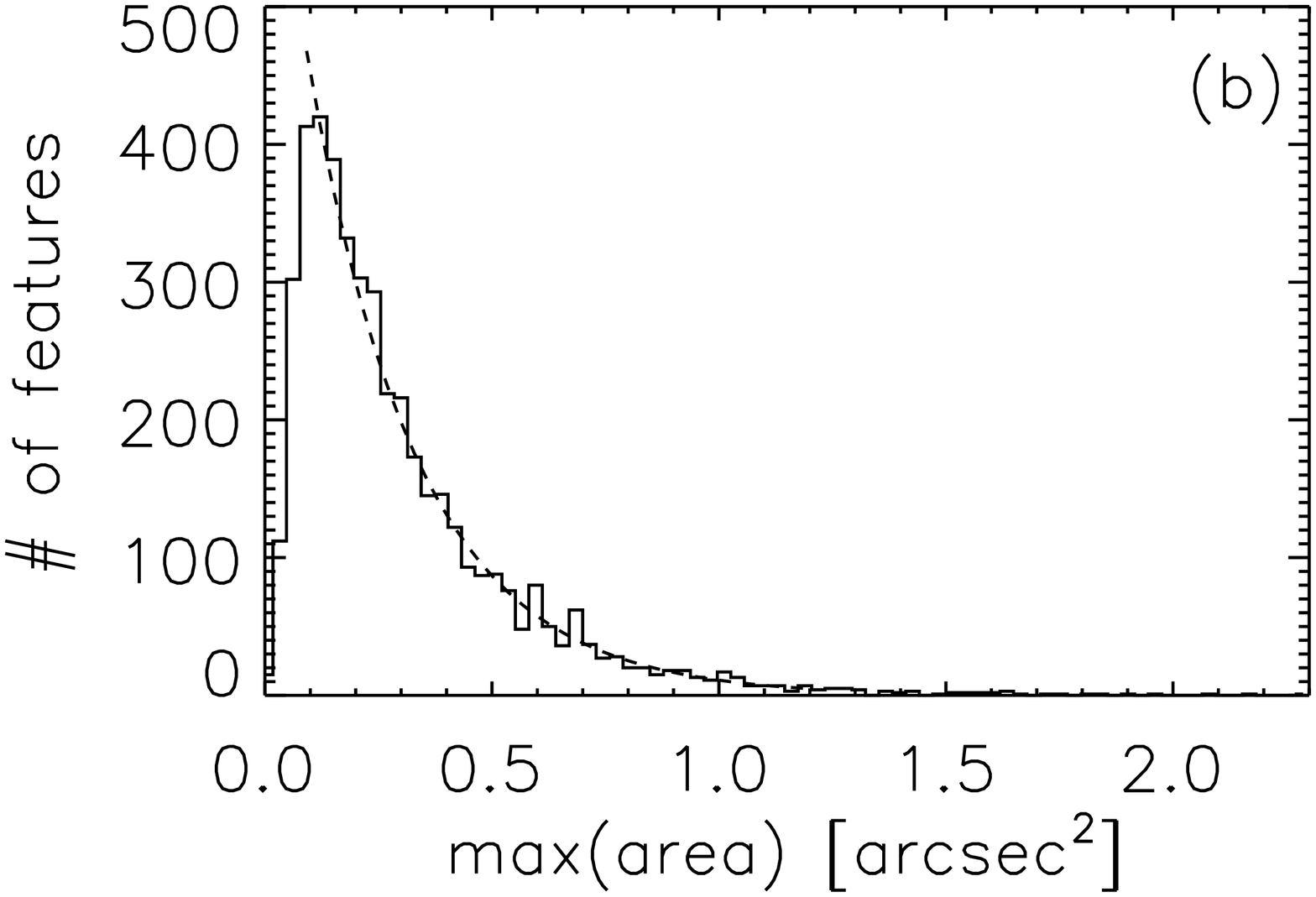}
\includegraphics*[angle=90,scale=.17]{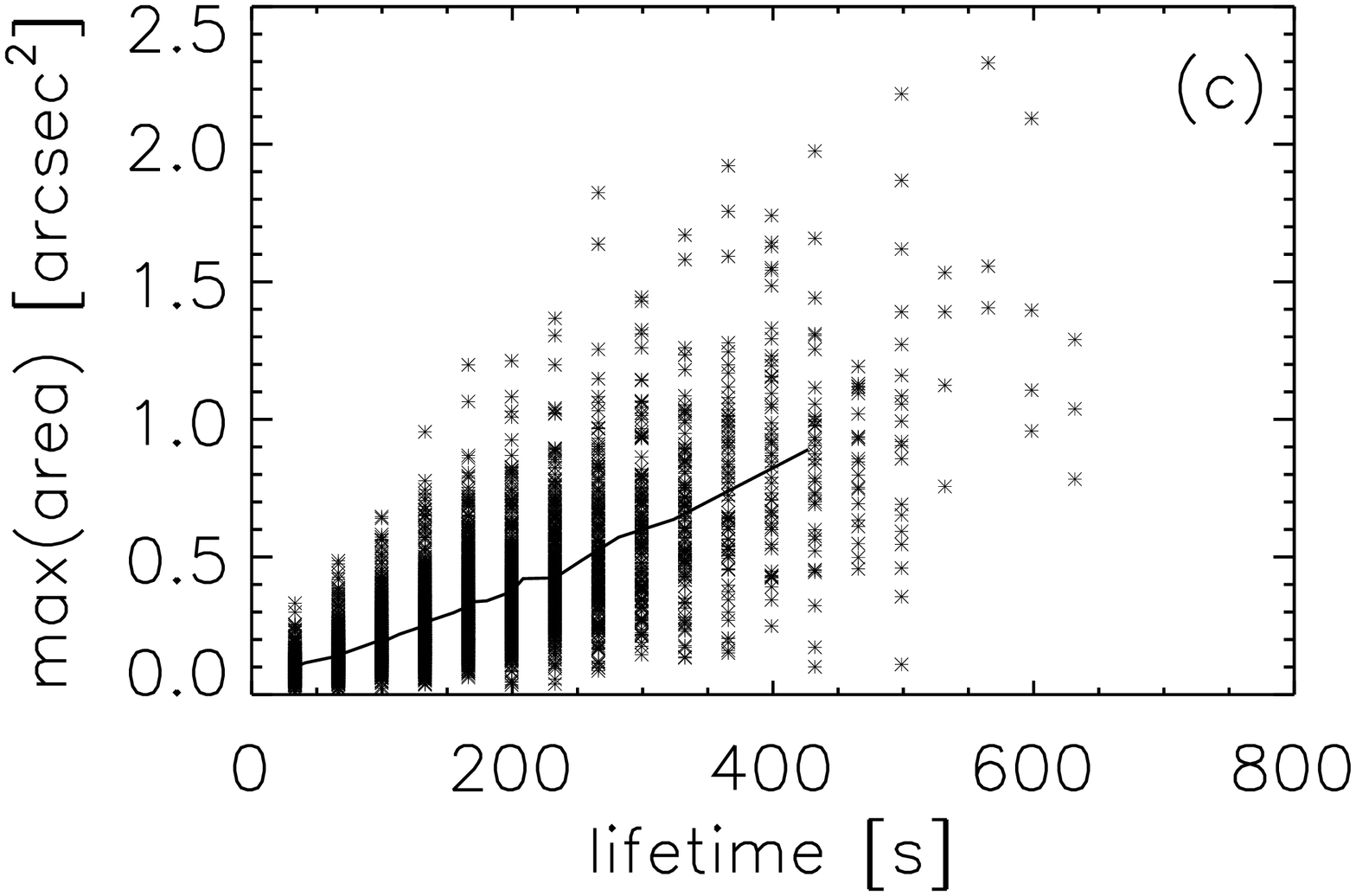}
\includegraphics*[angle=90,scale=.17]{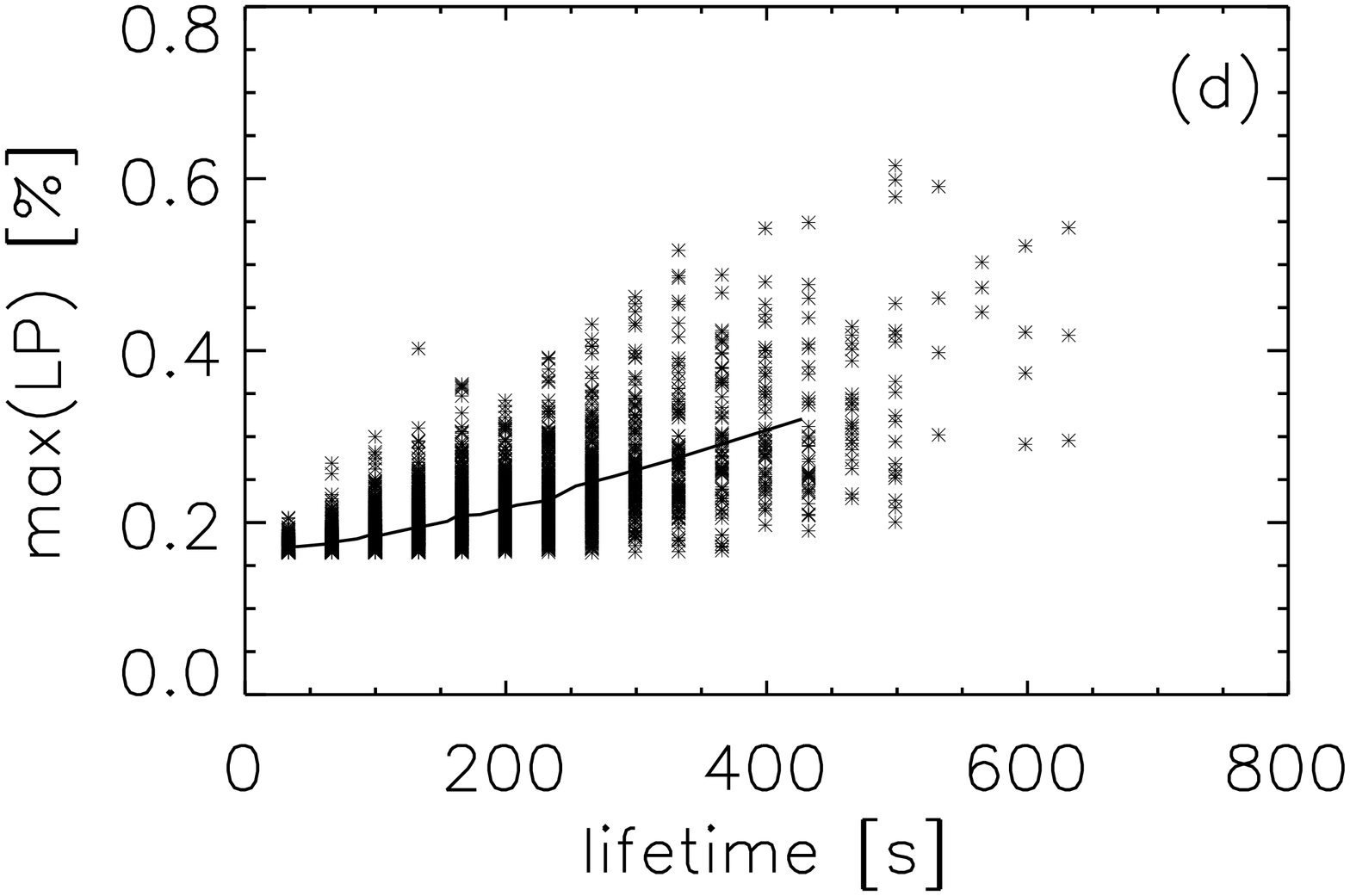}
\includegraphics*[angle=90,scale=.17]{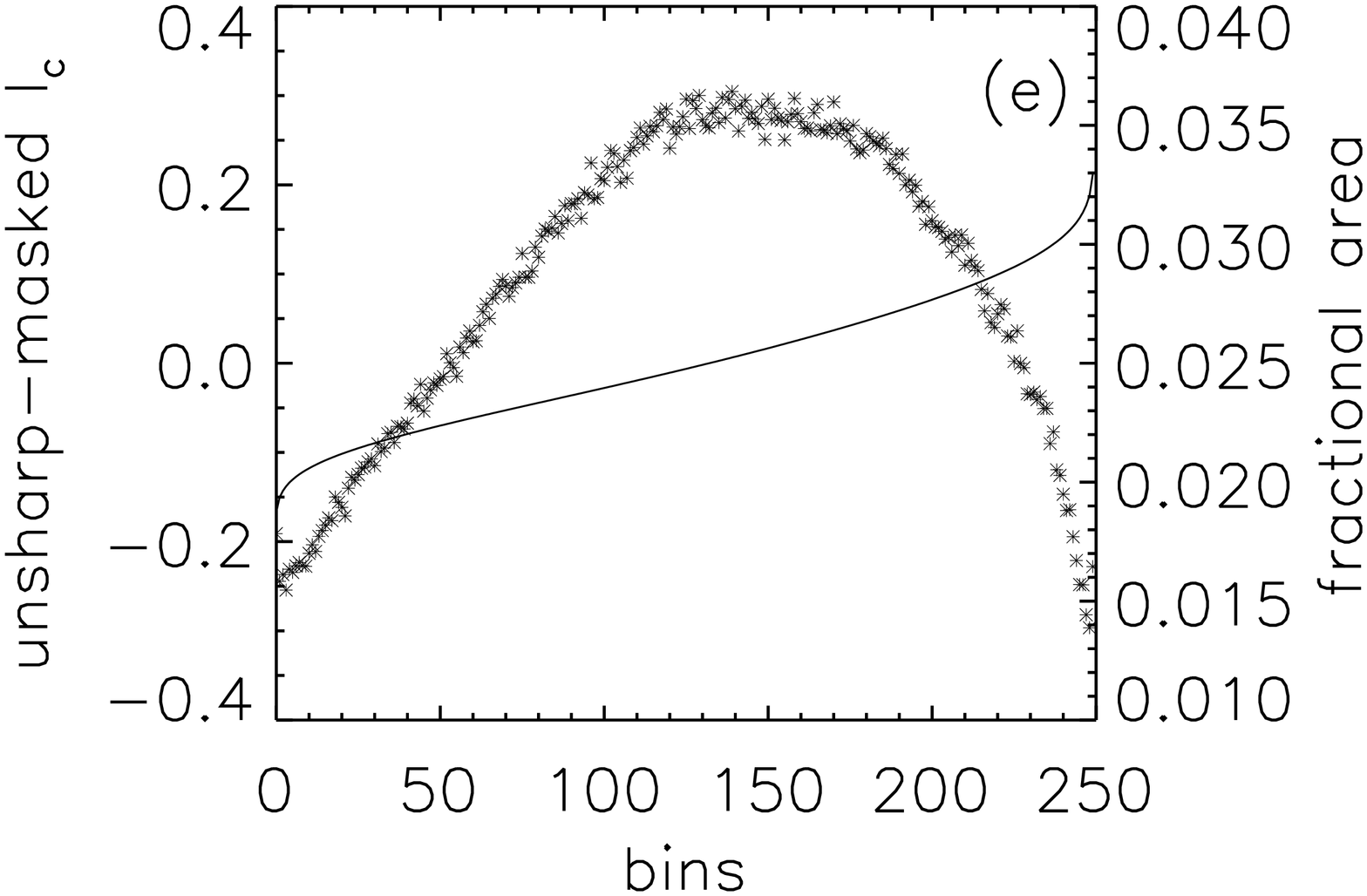}
\includegraphics*[angle=90,scale=.17]{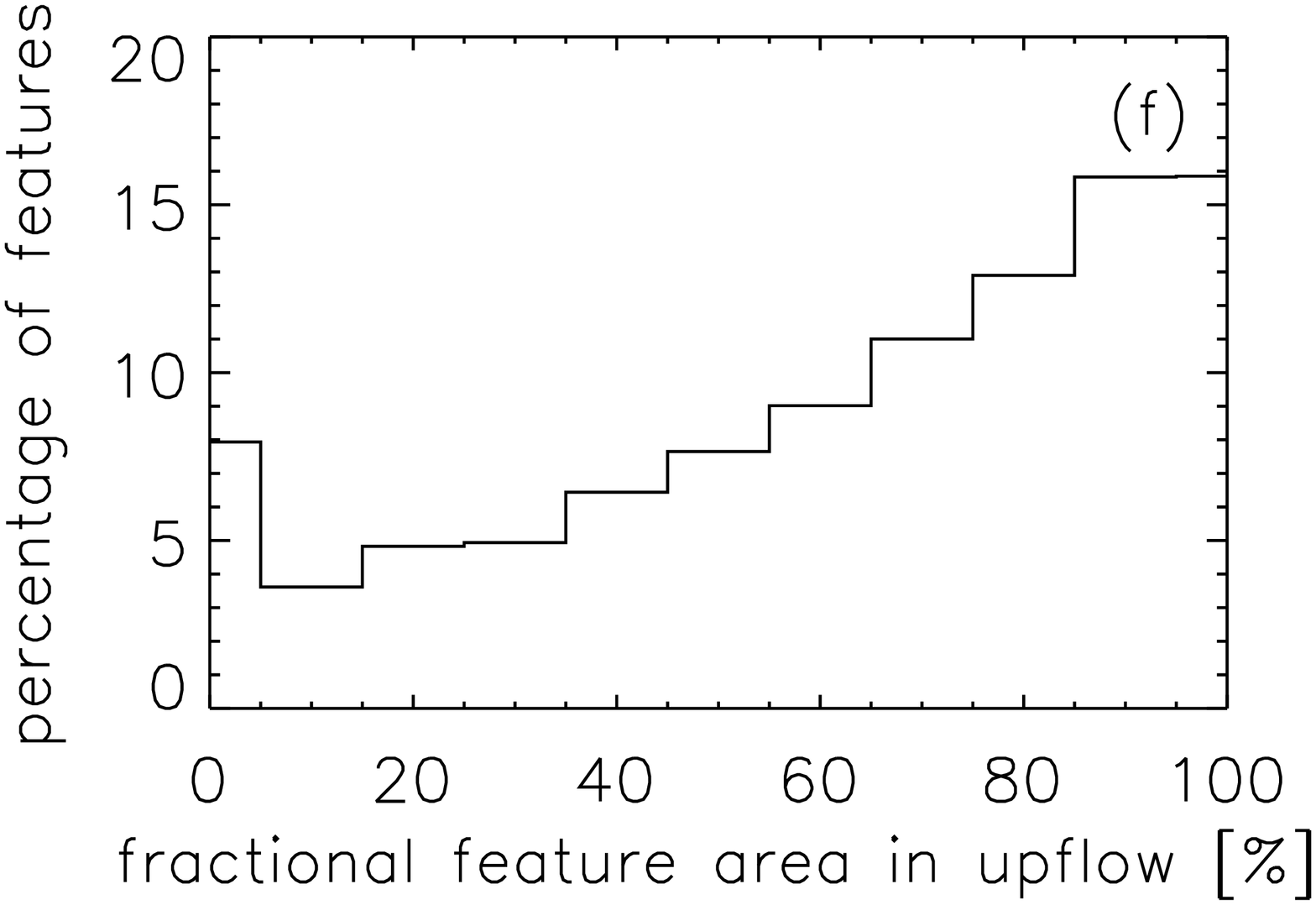}
\includegraphics*[angle=90,scale=.17]{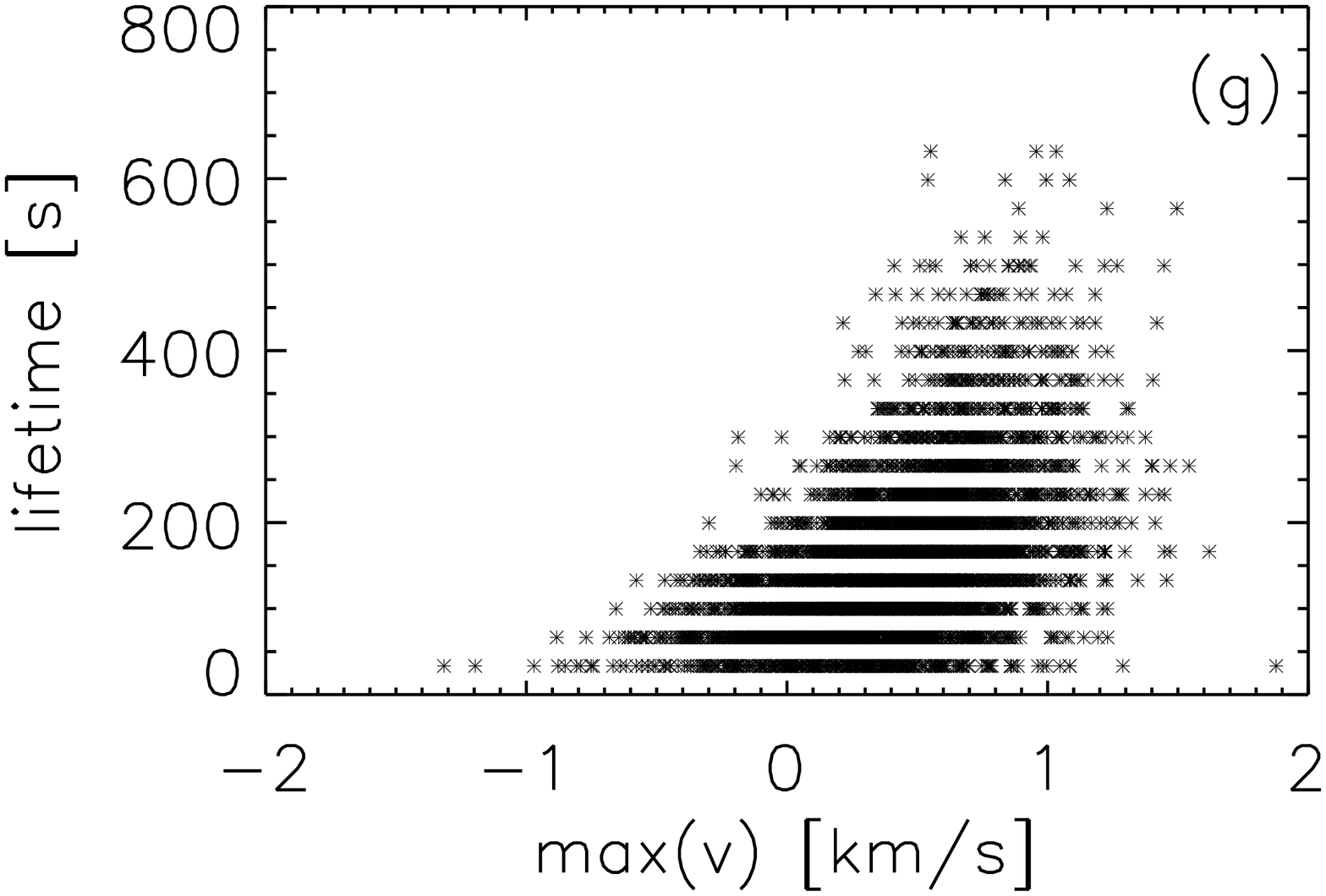}
\includegraphics*[angle=90,scale=.17]{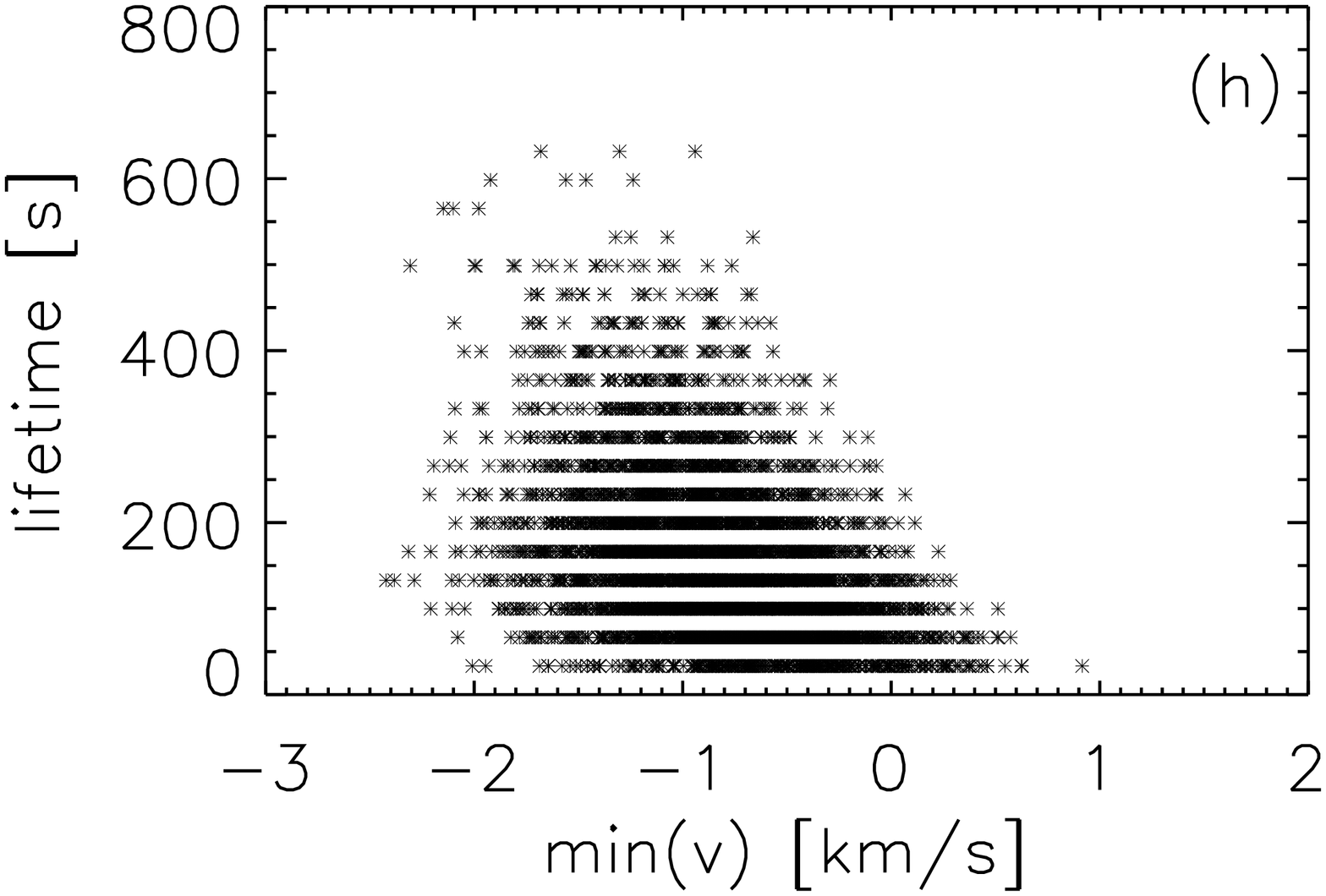}
\caption{Statistical properties of detected HIF. (a) Histogram of
lifetimes. The dashed line is an exponential fit with a decay time
of $92$~s. (b) Histogram of maximum area. The dashed line is an
exponential fit with a decay scale of $0.24$~arcsec$^{2}$. (c)
Scatter plot of maximum area as a function of lifetime. (d)
Scatter plot of maximum mean linear polarization signal as a
function of lifetime. (e) Position of the HIF with respect to
granules. The solid line indicates unsharp-masked
non-reconstructed continuum intensities divided into 250 equally
populated bins. The star symbols mark the fractional area of the
corresponding bins occupied by the HIF. (f) Histogram of the
percentage of the feature area coinciding with upflows, based on
non-reconstructed LOS velocities. (g) and (h) Scatter plots of the
feature lifetimes versus maximum and minimum velocity respectively
associated with the detected features. \label{fig:basics}}
\end{figure*}

\section{Observations}

We use two data sets obtained on June 9 2009, 00:36:02-00:58:46~UT
(data set 1) and 01:30:54-02:02:29~UT (data set 2). The FOV covers
$45^{\prime\prime}\times45^{\prime\prime}$ of a quiet region at
disk center. Polarization maps were taken in 5 wavelength
positions over the Fe~I~525.02~nm line with a cadence of $33$~s
and a pixel size of $0.055^{\prime\prime}$. After data reduction
\citep{Valentin:etal:2010}, two types of data are produced:
non-reconstructed (level 1) and data reconstructed using
phase-diversity information (level 2), reaching a spatial
resolution of 0.15-0.18 arcsec. All data are corrected for
instrumental effects, including intensity fluctuations due to
interference fringes, dust particles, and illumination effects, as
well as jitter-introduced polarization cross-talk, and blueshift
over the FOV due to the collimated setup of the magnetograph
etalon. The noise level of the non-reconstructed Stokes $Q$ and
$U$  data is $\sim10^{-3}I_c$. The reconstruction process
amplifies the power of all spatial frequencies and, therefore,
also increases the noise level by a factor of 2.5-3. Although the
polarization signals are also amplified by the reconstruction, a
significant amount is, nevertheless, lost in the noise. Thus, to
identify the HIF we use the non-reconstructed linear polarization
signal averaged over the 4 wavelength positions in the spectral
line \citep[][Eq. 15]{Valentin:etal:2010}. To reduce the noise
additionally, we applied the Gaussian filter with FWHM of 2 pixels
($0.11^{\prime\prime}$) to the linear polarization maps prior to
the HIF identification.

\section{Results}

To obtain statistical properties of a large number of features
with significant linear polarization signal, we use a modified
version of the MLT (Multi Level Tracking) algorithm of
\citet{Bovelet:Wiehr:2001} as an automatic detection method. MLT
applies a sequence of thresholds with decreasing values to the
mean linear polarization map. The algorithm identifies features
when they show the largest signal and expands them in three
dimensions (two spatial and one temporal) as the threshold value
is decreased. After extensive tests we chose to use 13 thresholds
ranging from $3.2\cdot10^{-3}I_c$ to a final threshold of
$1.5\cdot10^{-3}I_c$. To avoid artificial splitting of structures
with several intensity peaks we set an additional criterion. If
two features were separate for threshold $\textit{n}$ and their
maximum intensity does not exceed the $\textit{(n+1)}$th threshold
value by more than $17.5$\%, we let them merge
\citep{Bharti:etal:2010}. As an example, Fig.~\ref{fig:masks}
shows a polarization map and the result of its MLT segmentation.
The detected HIF have a wide range of sizes and appear to be
organized on mesogranular scales.
\begin{figure}
\center
\includegraphics[width=0.5\textwidth]{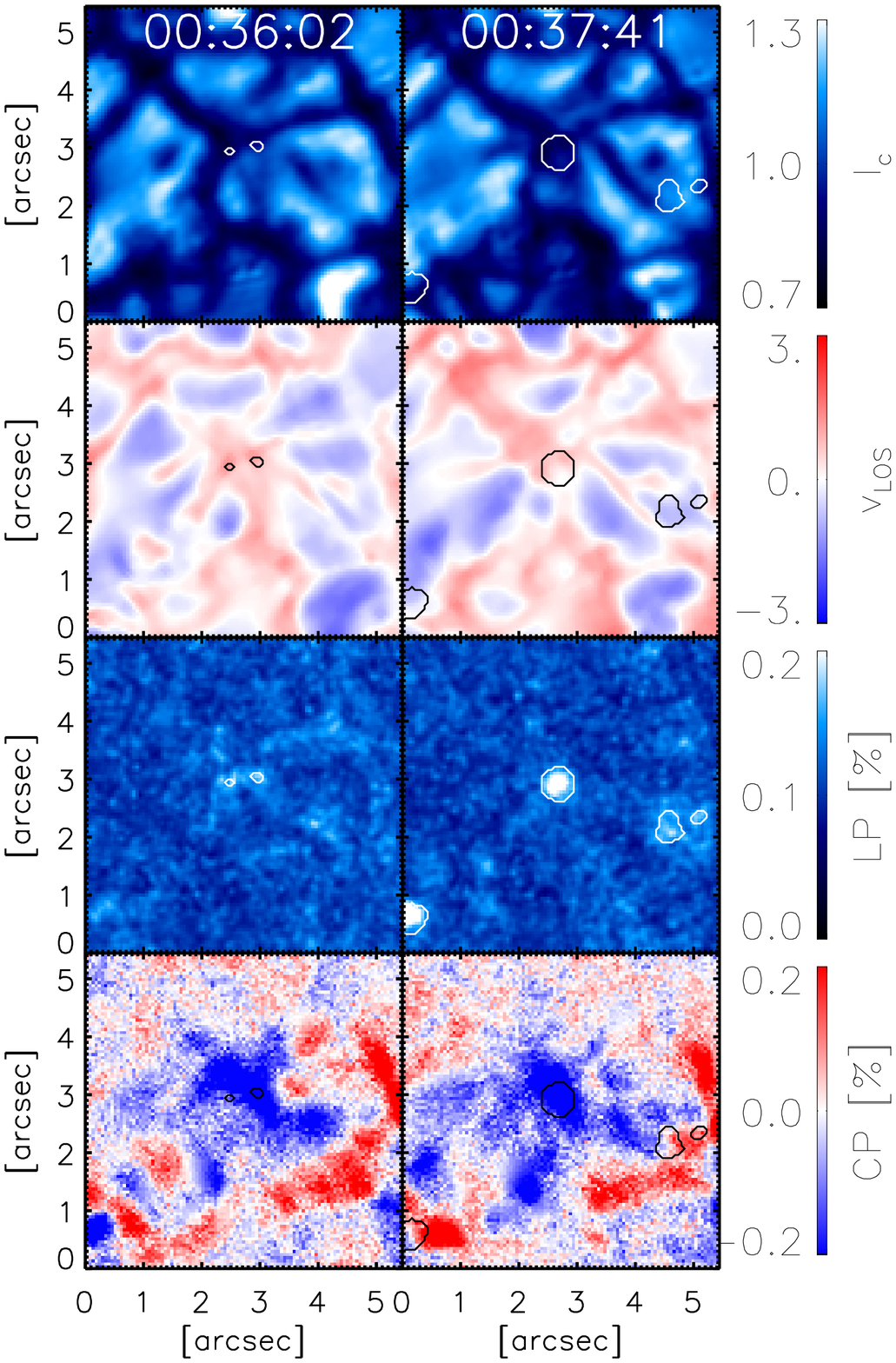}
\caption{Evolution of a HIF associated with a downflow (example
1). From top to bottom: continuum intensity, LOS velocity and maps
of the mean linear and circular polarization signal. The Gaussian
smoothing is applied to the linear polarization map
(FWHM=$0.11^{\prime\prime}$). Overplotted contours mark the
position of the features identified with MLT. The observations
were obtained at the times given in the top panels (in UT). The
HIF evolution during its whole lifetime (00:36-00:40~UT) is shown
in the movie provided as online material. \label{fig:ex1}}
\end{figure}

\subsection{Lifetime, size, and location}

The total number of features that are followed in time and space
from their appearance to their disappearance is 4536 (1911 in data
set 1 and 2625 in data set 2). Taking into account that the
features are detected during 22~min (data set 1) and 31~min (data
set 2) in a FOV of $45^{\prime\prime}\times45^{\prime\prime}$, we
obtain a rate of occurrence of
$7.1\cdot10^{-4}$~s$^{-1}$~arcsec$^{-2}$ (data set 1) and
$6.9\cdot10^{-4}$~s$^{-1}$~arcsec$^{-2}$ (data set 2). This is of
1-2 orders of magnitude higher than the values reported in
previous studies
\citep{Lites:etal:1996,Ishikawa:Tsuneta:2009,Jin:etal:2009,Marian:Luis:2009}.
The detected features occupy $\sim3$\% of the image area.

The distributions of lifetime and maximum area of the detected
features are shown in Fig.~\ref{fig:basics}a,b, respectively.
Since the duration of the observations is comparable to the
lifetimes of longer-lived features, we tend to underestimate their
number. Therefore, we correct the lifetime distribution by
multiplying with a weight of $(n-2)/(n-1-m)$ for structures that
live $m$ frames, where $n$ is the total number of frames. Owing to
the limited spatial resolution and the finite time cadence of our
observations, both distributions (lifetime and maximum area) show
peaks at small values. Both distribution also have the extended
tails that can be fitted with exponentials. This implies that the
features do not have a characteristic lifetime or size. Their
lifetimes range from $<$33~s (features that appear in only one
frame) to $10.5$~min, with $\sim$40\% of the features living less
than $100$~s. The area distribution has a peak at 0.1~arcsec$^2$,
with $\sim$12\% of the features being smaller than that. Around
$97$\% of the features are smaller than $\sim1$~arcsec$^2$, which
was given as the mean HIF size in previous studies
\citep{Ishikawa:Tsuneta:2009,Jin:etal:2009}.

Figures~\ref{fig:basics}c and d show scatter plots of maximum area
and maximum linear polarization signal versus feature lifetime.
Figure~\ref{fig:basics}d has a cutoff at $0.15$\%, which is the
lowest threshold value taken for MLT segmentation. The curves
connect binned values for $189$ points each. The plots indicate
that longer-lasting features tend to be larger, and to display a
higher mean linear polarization signal. The largest feature has a
lifetime of $9.4$~min and occupies up to $\sim2.3$~arcsec$^2$ in
the course of its evolution. Its mean linear polarization signal
reaches $0.5$\%.

In order to study whether the features are located in preferred
locations with respect to the granular pattern, we follow the
method of \citet{Lites:etal:2008}. Unsharp-masked continuum images
are obtained by subtracting, from the originals, the images
smoothed with a 59 pixel wide (3.2$^{\prime\prime}$) boxcar
function. In this way, intensity variations on scales larger than
granulation (due to, e.g., p-mode oscillations) are suppressed.
The pixels are then sorted into 250 equally populated intensity
bins, ranging from dark intergranuar lanes to bright granular
centers. The fractional area occupied by the HIF is calculated for
each bin. The results are shown in Fig.~\ref{fig:basics}e. The
solid line represents unsharp-masked continuum values (see y-axis
on the left). The right y-axis shows the range of fractional areas
occupied by the HIF in each bin. The distribution is similar to
the results shown by \citet[][Fig. 9]{Lites:etal:2008}. It has a
peak of 3.5\% at positive values of the unsharp-masked intensity
distribution. This suggests, as \citet{Lites:etal:2008} noticed,
that HIF tend to be located at intermediate intensities,
presumably at the periphery of granules.

\begin{figure}
\center
\includegraphics[width=0.5\textwidth]{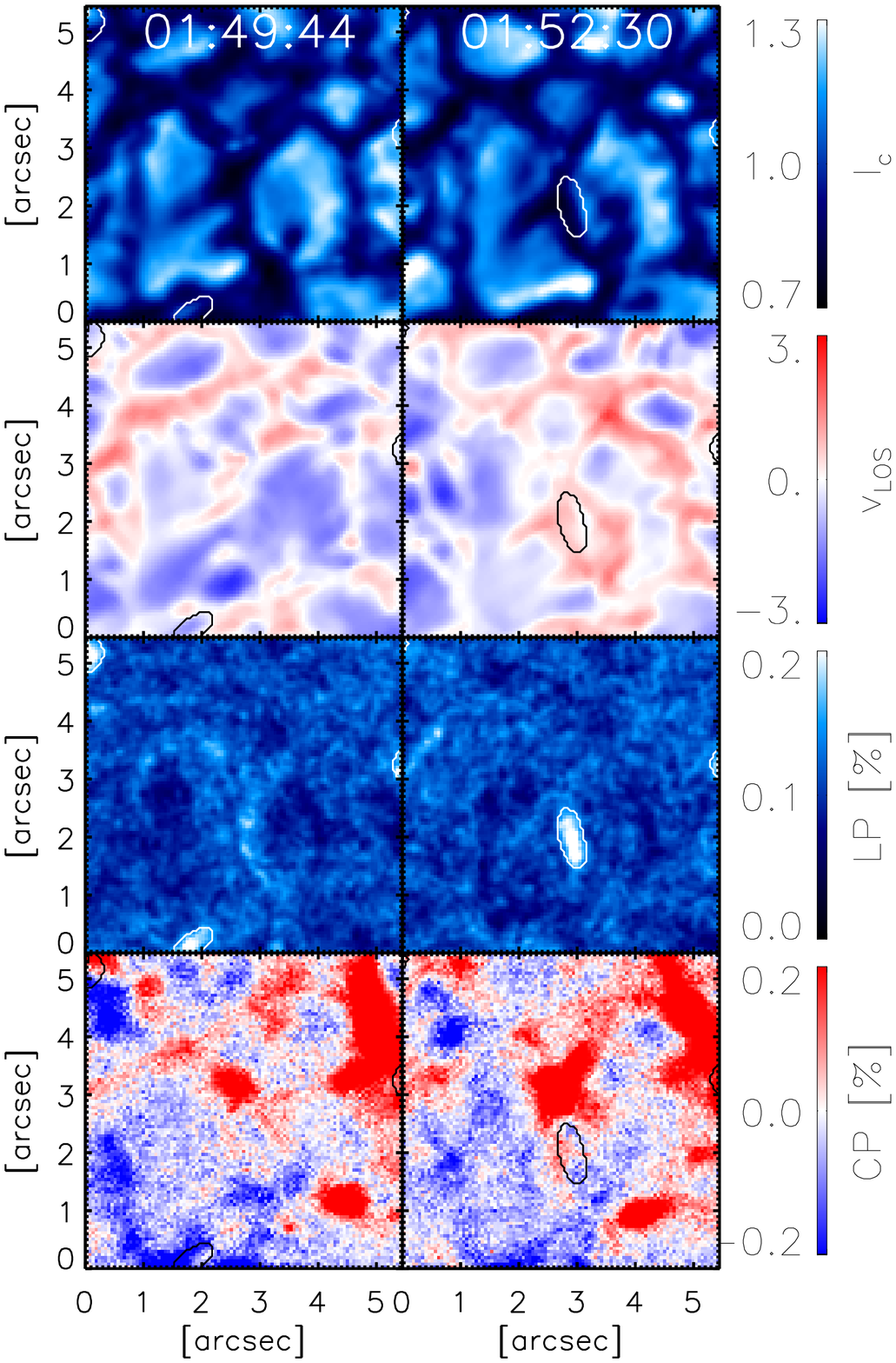}
\caption{Evolution of a HIF associated with a downflow (example
2). The format is the same as in Fig.~\ref{fig:ex1}. The HIF
evolution during its whole lifetime (01:49-01:55~UT) is shown in
the movie provided as online material. \label{fig:ex2}}
\end{figure}

\subsection{Emergence/submergence}

To estimate which percentage of the features are emerging or
submerging, we study the distribution of the associated line of
sight (LOS) velocities. The LOS velocities are derived from
Gaussian fits to the non-reconstructed Stokes~$I$ profiles.
Figure~\ref{fig:basics}f shows the fractional area of the detected
features associated with upflows. The distribution shows that the
majority of features have a large area fraction associated with
upflows. However, most of the features also have part of their
area associated with downflows. Only $\sim$16\% of the features
are fully embedded in upflows. The features fully embedded in
downflows make up $\sim$8\% of the total number of the detected
features.

Figures~\ref{fig:basics}g and ~\ref{fig:basics}h show HIF
lifetimes versus the minimum and maximum velocities, respectively,
associated with the HIF during their lifetimes. Positive
velocities correspond to downflows. The plots confirm that most
features are assocciated with both up- and downflows. Moreover,
the HIF tend to sample strong upflows and less strong downflows.
This implies that the linear polarization signal tends to
disappear before or when plasma overturns at granule edges. Only
$5$\% of the HIF show the stronger ($>1$~km/s) downflows in
intergranular lanes. Features that are fully embedded in upflows
(v$_{max}<0$) or downflows (v$_{min}>0$) during their whole
lifetime tend to be shortlived. They are also small in size
($<0.7$~arcsec$^2$ for the former and $<0.3$~arcsec$^2$ for the
latter).

Since emerging horizontal features have been studied in some
detail by other authors
\citep{Centeno:etal:2007,Marian:Luis:2009,Gomory:etal:2010}, we
consider here two examples of features associated with downflows.
Figures~\ref{fig:ex1} and ~\ref{fig:ex2} show the intensity
pattern and LOS velocity together with the mean linear and
circular polarization signals at two time instances: before and
during the HIF lifetime. Overplotted contours mark the position of
features detected with MLT. In Fig.~\ref{fig:ex1}, the continuum
images show a granule at
[4$^{\prime\prime}$,~2.5$^{\prime\prime}$] which is fragmenting
along two dark lanes. In the course of the fragmentation, one
downflow lane develops while another, already existing,
intergranular lane broadens. Adjacent to the granule, at the
junction of the lanes being formed, a patch of stronger linear
polarization signal appears at
[2.5$^{\prime\prime}$,~3$^{\prime\prime}$]. It disappears after
$2$~min, when the fragmentation process finishes. The circular
polarization maps show a dominant negative polarity patch at this
location. A small positive patch is also visible
$\sim1^{\prime\prime}$ away from the HIF (at
[3.5$^{\prime\prime}$,~3$^{\prime\prime}$]), at the beginning of
the HIF evolution. Towards the end, the magnetic concentration of
negative polarity also becomes weaker.

Figure~\ref{fig:ex2} presents a similar case. The granule at
[4$^{\prime\prime}$,~2$^{\prime\prime}$] is fragmenting followed
by the appearance of a strong downflow at the same location.
Linear polarization maps show a HIF appearing in the intergranular
lane, at [3$^{\prime\prime}$,~2$^{\prime\prime}$], growing with
time and then disappearing at the same location. During the same
period, two adjacent patches of opposite polarity are visible in
the circular polarization maps. The positive polarity patch
increases in size, while the negative becomes weaker at the
location of the downflow (at
[3$^{\prime\prime}$,~1$^{\prime\prime}$]).

The example shown in Fig.~\ref{fig:ex2} belongs to $\sim$53\% of
the detected features that have circular polarization signal (CP)
of both polarities associated with them. The features that are
associated with only one polarity (as the example shown in
Fig.~\ref{fig:ex1}) make up $\sim$42\% of the total number of the
detected features. The rest of the HIF has no CP higher than
$2\sigma$ ($0.2$\%) in their vicinity (in the region of
$\leq3$~pixels around them). If we consider only features that are
fully embedded in downflows, then $\sim$60\% of them are
associated with only one polarity and $\sim$31\% is associated
with CP of both polarities.

Considering results from MHD simulations, we can think of two
possible scenarios for horizontal fields associated with
downflows. In the first, a magnetic loop is submerged by a
downflow \citep{Stein:Nordlund:2006}. In the second scenario, the
field is organized in small loop-like structures and forms bundles
which, observed with limited spatial resolution, appear as patches
of higher linear polarization signal located in downflow lanes
\citep{Danilovic:etal:2010}. As flux is being redistributed owing
to the granular evolution, the bundles are dispersed and the
spatial smearing of more isolated loop-like structures reduces the
linear polarization signal to values below the noise level.

\section{Summary}

Based on Sunrise/IMaX data and using an automated detection
method, we obtained statistical properties of 4536 features with
significant linear polarization signal. Their lifetimes are
consistent with examples given previously in the literature.
However, the lifetime distribution indicates no characteristic
value, in contrast to previous studies
\citep{Ishikawa:etal:2008,Ishikawa:Tsuneta:2009,Jin:etal:2009}.
The detected features have no characteristic size either. Around
$97$\% of them are smaller than $\sim$1~arcsec$^2$, which is the
value previously taken as the mean size of HIF
\citep{Ishikawa:Tsuneta:2009}. We find that their rate of
occurrence is 1-2 orders of magnitude higher than reported earlier
\citep{Lites:etal:1996,Ishikawa:Tsuneta:2009,Marian:Luis:2009}. We
attribute this discrepancy to selection effects. If we take only
the biggest features (sizes $>0.88$~arcsec$^2$), only $\sim$4\% of
the detected features remain and the rate of occurrence decreases
to $\sim4\cdot10^{-5}$~s$^{-1}$~arcsec$^{-2}$, which is in closer
agreement with the references cited above. Longer-lived HIF tend
to be larger and display a higher mean linear polarization
signals. The HIF appear preferentially at the granule boundaries,
with most of them being caught by downflows at some point in their
evolution. We showed that $\sim$16\% of the features we detected
are completely embedded in upflows and $\sim$8\% are entirely
embedded in downflows. The latter are very small in size (as
illustrated by the two examples discussed in greater detail).
Although their origin is still uncertain it is clear that they do
not fit into the scenario of magnetic flux emergence as their
physical cause.

  \begin{acknowledgements}
   The German contribution to $Sunrise$ is funded by the Bundesministerium
   f\"{u}r Wirtschaft und Technologie through Deutsches Zentrum f\"{u}r Luft-
   und Raumfahrt e.V. (DLR), Grant No. 50~OU~0401, and by the Innovationsfond of
   the President of the Max Planck Society (MPG). The Spanish contribution has
   been funded by the Spanish MICINN under projects ESP2006-13030-C06 and
   AYA2009-14105-C06 (including European FEDER funds). The HAO contribution was
   partly funded through NASA grant number NNX08AH38G. This work
   has been partially supported by the WCU grant No. R31-10016
   funded by the Korean Ministry of Education, Science and
   Technology.
   \end{acknowledgements}

\end{document}